\documentclass[final,5p,times,twocolumn]{elsarticle}
\usepackage{lineno,hyperref}
\usepackage{amsmath,amssymb}
\usepackage{amsfonts,amstext}
\usepackage{epstopdf}
\usepackage{enumitem}
\usepackage{psfrag}
\usepackage{array}
\usepackage{epstopdf}
\usepackage{booktabs}
\usepackage{verbatim}
\usepackage{cases}
\usepackage{MnSymbol}
\usepackage{float}
\usepackage{longtable}
\usepackage[center]{caption}
\allowdisplaybreaks
\newcommand{\vecnm}[1]{\left\|#1\right\|}
\usepackage{graphicx,tikz,pgfplots}
\usepackage{color}
\usepackage{soul}

\usepackage{amssymb}
\usepackage{amsthm}
\usepackage{cite}

\usepackage[utf8]{inputenc} 
\usepackage{titlesec}
\titleformat{\paragraph}
{\normalfont\normalsize\bfseries}{\theparagraph}{1em}{}
\titlespacing*{\paragraph}
{0pt}{3.25ex plus 1ex minus .2ex}{1.5ex plus .2ex}
\usepackage[final]{epsfig}
\usepackage{graphics}
\usepackage{float}
\usepackage{array}
\usepackage{amsmath,amssymb,latexsym}
\usepackage{multirow}

\newcommand*{\bm}[1]{{\color{black} #1}}
\newcommand*{\nl}[1]{{\color{black} #1}}

\begin{document}

\begin{frontmatter}
\author[label1b]{Bruno M\'{e}riaux\fnref{t1,t2}}
\fntext[t1]{The authors have equal contribution to the paper.}
\fntext[t2]{This work is partially funded by the Direction G\'{e}n\'{e}rale de l'Armement (D.G.A.) as well as the ANR ASTRID referenced ANR-17-ASTR-0015.}
\author[label1]{Xin Zhang\fnref{t1}}
\author[label2]{Mohammed Nabil El Korso\corref{cor1}}
\ead{m.elkorso@parisnanterre.fr}
\cortext[cor1]{Corresponding author}
\author[label3]{Marius Pesavento}
\address[label1b]{SONDRA, CentraleSup\'{e}lec, 91192, Gif-sur-Yvette, France}
\address[label1]{AI Department, Echiev Autonomous Driving Technology, 518057, Shenzhen, China}
\address[label2]{Paris Nanterre University, LEME EA-4416 92410 Ville d'Avray, France}
\address[label3]{Communication Systems Group, Technische Universit\"{a}t Darmstadt, Darmstadt 64283, Germany}

\title{Iterative Marginal Maximum Likelihood DOD and DOA Estimation for MIMO Radar in the Presence of SIRP Clutter}

\begin{abstract}
The spherically invariant random process (SIRP) clutter model is commonly used in scenarios where the radar clutter cannot be correctly modeled as a Gaussian process. In this short communication, we devise a novel Maximum-Likelihood (ML)-based iterative estimator for direction-of-departure and direction-of-arrival estimation in the Multiple-input multiple-output (MIMO) radar context in the presence of SIRP clutter. The proposed estimator employs a stepwise numerical concentration approach w.r.t. the objective function related to the marginal likelihood of the observation data. Our estimator leads to superior performance, as our simulations show, w.r.t. to the existing likelihood based methods, namely, the conventional, the conditional and the joint likelihood based estimators, and w.r.t. the robust subspace decomposition based methods. Finally, interconnections and comparison between the Iterative Marginal ML Estimator (IMMLE), Iterative Joint ML Estimator (IJMLE) and Iterative Conditional ML Estimator (ICdMLE) are provided.
\end{abstract}
\end{frontmatter}
\section{Introduction}
Multiple-input multiple-output (MIMO) radar has found wide application in the past decades. By means of waveform diversity, MIMO radar allows significant improvement of performance to be made as compared with the conventional phased-array radar \citep{LS08}. There exist in the literature abundant works to investigate algorithms for target localization or to evaluate their performances in MIMO radar contexts \citep{JLL09,hong1,tang1} mostly under the umbrella of Gaussian clutter. The validity of the Gaussian clutter assumption is rooted in the central limit theorem and is realistic in the case of sufficiently large number of independent and identically distributed (i.i.d.) elementary scatterers. In applications of high-resolution radars, the radar clutter exhibits non-stationarity, and a Gaussian modeling of the clutter, be it white or colored, deviates heavily from the real data and thus is inadequate \citep{mim5}.

\nl{Non-Gaussian clutter scenarios have been first studied through $\alpha$-stable distribution and  mixture noise distributions. Nevertheless, the so-called spherically invariant random process (SIRP) has, thanks to its ability to describe different scales of the clutter roughness and to incorporate various non-Gaussian distributions, become a favorite distribution family in the radar context \citep{mim5,mim12}. The latter is a two-scale, compound Gaussian process which is a product of two components: the {texture} and the {speckle}. The texture, which accounts for \bm{local power changes}, is the square root of a positive scalar random process, whereas the speckle,  which accounts for a local scattering, is a complex Gaussian process.
Though abounding works have been dedicated to the estimation algorithms in the SIRP clutter context with zero mean observations \citep{mim14,mim15,mim17,gini1}, there are, to the best of our knowledge, few works dealing with jointly parameterized mean and parameterized covariance matrix in the context of SIRP clutter \citep{xzh4,xzh5,BAJ16}. Among these few works, we can, \bm{first}, cite the robust MUSIC (MUSIC-Tyler) based on a robust fixed point Tyler estimate of the covariance matrix \citep{Tyl87}, the robust covariation-based MUSIC (ROC-MUSIC) \citep{TN96}, adapted for $\alpha$-stable distribution, in which MUSIC method is applied to the covariation matrix instead of the estimated covariance matrix and the RG-MUSIC \citep{Cou15} based on the random matrix theory (namely, it takes into account the Marcenko-Pastur distribution of the eigenvalues of the covariance matrix to rectify its estimation). \bm{On the other hand,} the $\ell_{p}$-MUSIC \citep{ZSH13} \bm{is} based on $\ell_{p}$ norm minimization with $p<2$ in order to take into account impulsive noise. \bm{Finally, some algorithms rely on robust mixtures noise} as \citep{BAJ16,KS00}, in which the authors proposed respectively, ML based method in the presence of a mixture of K-distributed and Gaussian noise \citep{BAJ16} and ML based method in non-Gaussian noise with Gaussian mixtures \citep{KS00}.}

In this short communication, we focus on the  direction-of-departure (DOD) and/or direction-of-arrival (DOA) estimation problems in the presence of SIRP noise/clutter, under an array processing model and a MIMO radar model. In \citep{xzh4,xzh5}, the authors designed estimators based on the Iterative Conditional Maximum Likelihood Estimator (ICdMLE) and the Iterative Joint Maximum Likelihood Estimator (IJMLE), which are, based on the \emph{conditional} likelihood of the observations on the texture realizations, and the \emph{joint} likelihood between the two, respectively. As a consequence, these two estimators are both \textit{eo ipso} suboptimal.

To overcome the algorithm suboptimality and the model limitations in \citep{xzh4,xzh5} (as the existence of only one Coherent Pulse Interval (CPI), the fact that DOD and DOA are assumed to share same values), we propose in this short communication an iterative ML estimator that is based on the marginal (\emph{exact}) observation likelihood for a general MIMO radar model under SIRP clutter, named the Iterative Marginal ML Estimator (IMMLE). As our derivations will show, the MIMO model in this paper after matched-filtering can be transformed into the same structure as the array processing model considered in \citep{xzh4}, meaning that the proposed IMMLE is directly applicable to the latter model without any further generalization.

\section{Model setup}\label{sec2}

\subsection{Observation model}

Consider a MIMO radar system with linear and possibly non-uniform arrays both at the transmitter and the receiver. Further assume that $K$ targets are illuminated by the MIMO radar, all modeled as far-field, narrowband, point sources \citep{LS08}. The radar output for the \textit{l}th pulse in a CPI, and after matched filtering in the case of transmission of orthogonal waveforms \citep{HBC08}, reads:
\begin{align}
\boldsymbol{Z}(l)=\frac{1}{\sqrt{T}}\boldsymbol{Y}(l)\boldsymbol{S}^H
&=\sum_{k=1}^{K}\sqrt{T}\alpha_{k}e^{2j\pi f_k l}\boldsymbol{a}_{(\mathcal R)}\left(\theta^{(\mathcal R)}_{k}\right)\boldsymbol{a}_{(\mathcal T)}^{T}\left(\theta^{(\mathcal T)}_{k}\right)\nonumber\\
&\quad +\boldsymbol{N}(l), \ \ \ \mbox{for} \ \ \ l=0,\dots,L-1
\label{1c}
\end{align}
where $L$ denotes the number of radar pulses per CPI; $\alpha_{k}$ and $f_k$ denote a complex coefficient proportional to the radar cross section (RCS) and the normalized Doppler frequency of the \textit{k}th target, respectively; $T$ is the number of snapshots per pulse, $\theta^{(\mathcal T)}_{k}$ and $\theta^{(\mathcal R)}_{k}$ represent the DOD and DOA of the \textit{k}th target, respectively; the transmit and receive steering vectors are defined as
$\boldsymbol{a}_{(\mathcal T)}(\theta^{(\mathcal T)}_{k})=[e^{j\frac{2\pi \sin\left(\theta^{(\mathcal T)}_{k}\right)}{\lambda}d_{1}^{(\mathcal T)}},\dots, e^{j\frac{2\pi \sin\left(\theta^{(\mathcal T)}_{k}\right)}{\lambda}d_{M}^{(\mathcal T)}}]^{T}$ and $\boldsymbol{a}_{(\mathcal R)}(\theta^{(\mathcal R)}_{k})=[e^{j\frac{2\pi \sin\left(\theta^{(\mathcal R)}_{k}\right)}{\lambda}d_{1}^{(\mathcal R)}},\dots, e^{j\frac{2\pi \sin\left(\theta^{(\mathcal R)}_{k}\right)}{\lambda}d_{N}^{(\mathcal R)}}]^{T}$, in which $M$ and $N$ represent the number of sensors at the transmitter and the receiver, respectively; $d_{i}^{(\mathcal T)}$ and $d_{i}^{(\mathcal R)}$ denote the distance between the $i$th sensor and the reference sensor for the transmitter and the receiver, respectively; $\lambda$ stands for the wavelength; $\boldsymbol{N}(l)$ denotes the received clutter matrix at pulse $l$; and $(\cdot)^{T}$ denotes the transpose of a matrix.

By stacking the output in Eq.~(\ref{1c}) into an $MN\times1$ vector denoted by $\boldsymbol{z}(l)$, we further have:
\begin{equation}
\boldsymbol{z}(l)=\text{vec}\left\{\boldsymbol{Z}(l)\right\}=\boldsymbol{A}\left(\boldsymbol{\theta}\right)\boldsymbol{v}(l)
+\boldsymbol{n}(l),\quad l=0,\dots,L-1,
\end{equation}in which
$\boldsymbol{A}\left(\boldsymbol{\theta}\right)=\left[\boldsymbol{a}\left(\theta^{(\mathcal T)}_{1},\theta^{(\mathcal R)}_{1}\right),\dots,\boldsymbol{a}\left(\theta^{(\mathcal T)}_{K},\theta^{(\mathcal R)}_{K}\right)\right]$ denotes the steering matrix after matched filtering, where $\boldsymbol{\theta}=\left[\theta^{(\mathcal T)}_{1},\theta^{(\mathcal R)}_{1},\dots,\theta^{(\mathcal R)}_{K}\right]^T$ is a vector parameter introduced to incorporate all the unknown DODs and DOAs of the targets, and, $
\boldsymbol{a}\left(\theta^{(\mathcal T)}_{k},\theta^{(\mathcal R)}_{k}\right)=\text{vec}\left\{\boldsymbol{a}_{(\mathcal R)}\left(\theta^{(\mathcal R)}_{k}\right)\boldsymbol{a}_{(\mathcal T)}^{T}\left(\theta^{(\mathcal T)}_{k}\right)\right\}
=\left(\boldsymbol{I}_M\otimes\boldsymbol{a}_{(\mathcal R)}\left(\theta^{(\mathcal R)}_{k}\right)\right)\boldsymbol{a}_{(\mathcal T)}\left(\theta^{(\mathcal T)}_{k}\right),$ in which $\boldsymbol{I}_M$ stands for the identity matrix of size $M$, and $\otimes$ denotes the Kronecker product; $\boldsymbol{v}(l)=\left[\sqrt{T}\alpha_{1}e^{2j\pi f_1 l},\dots,\sqrt{T}\alpha_{K}e^{2j\pi f_K l}\right]^T$; $\boldsymbol{n}(l)=\text{vec}\left\{\boldsymbol{N}(l)\right\}$ denotes the clutter vector after matched filtering at pulse $l$; and $\text{vec}\{\cdot\}$ stands for the vectorization of a matrix.

\subsection{Observation statistics}\label{IIB}
We model the clutter vectors $\boldsymbol{n}(l),\ l=0,\dots,L-1$ as independent, identically distributed (i.i.d.) Spherically Invariant Random Vectors (SIRVs), which can be formulated as the product of two components statistically independent of each other:
$\boldsymbol{n}(l)=\sqrt{\tau(l)}\boldsymbol{x}(l),\,l=0,\dots, L-1$, in which the texture terms $\tau(l)$, are i.i.d. positive random variables; the speckle terms $\boldsymbol{x}(l)$ are i.i.d. $MN$-dimensional circular complex Gaussian vectors with zero mean and second-order moments $\text{E}\left\{\boldsymbol{x}(i)\boldsymbol{x}^H(j)\right\}=\delta_{ij}\boldsymbol{\Sigma}$ where $\boldsymbol{\Sigma}$ denotes the speckle covariance matrix, $\text{E}{\{\cdot\}}$ is the expectation operator, $\delta_{ij}$ is the Kronecker delta. To avoid the ambiguity in the model arising from the scaling effect between the texture and the speckle, we assume that $\text{tr}\{\boldsymbol{\Sigma}\}=MN$, in which $\text{tr}\{\cdot\}$ denotes the trace.
In this paper, we mainly focus on two kinds of SIRP clutters that are prevalent in the literature, namely, the K-distributed and the t-distributed clutters. In both cases the texture is characterized by two parameters, the \emph{shape parameter} $a$ and the \emph{scale parameter} $b$:
\begin{itemize}
\item \textbf{K-distributed clutter}, in which $\tau(l)$ follows a \emph{gamma distribution} (denoted by $\tau(l)\sim \text{Gamma}(a, b)$), namely, $p(\tau(l);a,b)=\frac{1}{\Gamma(a)b^a}\tau(l)^{a-1}e^{-\frac{\tau(l)}{b}},$
in which $\Gamma(\cdot)$ denotes the gamma function.
\item \textbf{t-distributed clutter}, in which $\tau(l)$ follows an \emph{inverse-gamma distribution} (denoted by $\tau(l)\sim \text{Inv-Gamma}(a, b)$), thus, $
p(\tau(l);a,b)=\frac{b^a}{\Gamma(a)}\tau(l)^{-a-1}e^{-\frac{b}{\tau(l)}}.$
\end{itemize}

\subsection{Unknown parameter vector and likelihood function}
Under the assumptions above, the unknown parameter vector of our problem is given by:
\begin{equation}\label{xi}
\boldsymbol{\xi}=\left[\boldsymbol{\theta}^T,\Re\left\{\boldsymbol{\alpha}\right\}^T,
\Im\left\{\boldsymbol{\alpha}\right\}^T,\boldsymbol{f}^T,\boldsymbol{\zeta}^T,a,b\right]^T,
\end{equation}in which $\boldsymbol{\alpha}=\left[\alpha_1,\dots,\alpha_K\right]^T$ is a complex vector parameter including the RCS coefficients of all $K$ targets, $\boldsymbol{f}=\left[f_1,\dots,f_K\right]^T$ contains the normalized Doppler frequencies of the targets, $\boldsymbol{\zeta}$ is a $M^2N^2$-element vector containing the real and imaginary parts of the entries of the lower triangular part of $\boldsymbol{\Sigma}$, $\Re\{\cdot\}$ and $\Im\{\cdot\}$ denote the real and the imaginary part, respectively.

Let $\boldsymbol{z}=\left[\boldsymbol{z}^T(1),...,\boldsymbol{z}^T(L-1)\right]^T$ denotes the full observation vector after matched filtering, and $\boldsymbol{\tau}=\left[\tau(0),\dots,\tau(L-1)\right]^T$ represents the vector of texture realizations at all pulses. The full observation likelihood conditioned on $\boldsymbol{\tau}$ can be written as:
\begin{equation}\label{n1}
p\left(\boldsymbol{z} \left\mid \boldsymbol{\tau}; \boldsymbol{\bar{\xi}}\right. \right)=\prod_{l=0}^{L-1}\frac{\exp\left(-\frac{\vecnm{\boldsymbol{\rho}(l)}^2}{\tau(l)}\right)}
{\left|\pi\boldsymbol{\Sigma}\right|\tau^{MN}(l)};
\end{equation}in which $\boldsymbol{\bar{\xi}}=\left[\boldsymbol{\theta}^T,\Re\left\{\boldsymbol{\alpha}\right\}^T,
\Im\left\{\boldsymbol{\alpha}\right\}^T,\boldsymbol{f}^T,\boldsymbol{\zeta}^T\right]^T$ is the unknown parameter vector that does not contain the texture parameters $a$ and $b$, $\vecnm{\cdot}$ denotes the norm of a vector, and
\begin{equation}\label{rho}
\boldsymbol{\rho}(l)=
\boldsymbol{\Sigma}^{-1/2}\left(\boldsymbol{z}(l)-\boldsymbol{A}\left(\boldsymbol{\theta}\right)\boldsymbol{v}(l)\right),
\end{equation} which represents the clutter realization at pulse $l$ with its speckle spatially whitened.
 The conditional likelihood in Eq.~(\ref{n1}), multiplied by $p(\boldsymbol \tau;a,b)$, leads to the joint likelihood between $\boldsymbol{z}$ and $\boldsymbol{\tau}$:
{\small \begin{equation}\label{n1a}
p\left(\boldsymbol{z}, \boldsymbol{\tau}; \boldsymbol{\xi} \right)
=p\left(\boldsymbol{z} \left\mid \boldsymbol{\tau}; \boldsymbol{\bar{\xi}}\right.\right)p(\boldsymbol \tau;a,b)
=\prod_{l=0}^{L-1}\frac{\exp\left(-\frac{\vecnm{\boldsymbol{\rho}(l)}^2}{\tau(l)}
\right)}{\left|\pi\boldsymbol{\Sigma}\right|\tau^{MN}(l)}p(\tau(l);a,b).
\end{equation}}

Finally, the full observation marginal (exact) likelihood, w.r.t. $\boldsymbol{\xi}$, is obtained by integrating out $\boldsymbol{\tau}$ from the joint likelihood in Eq.~(\ref{n1a}), as:
\begin{align}
p\left(\boldsymbol z; \boldsymbol{\xi} \right)=
\int_0^{+\infty}p\left(\boldsymbol z, \boldsymbol{\tau}; \boldsymbol{\xi} \right)\text{d}{\boldsymbol \tau}
&=\prod_{l=0}^{L-1}\int_{0}^{+\infty}\frac{\exp\left(-\frac{\vecnm{\boldsymbol{\rho}(l)}^2}
{\tau(l)}\right)}{\left|\pi\boldsymbol{\Sigma}\right|\tau^{MN}(l)} \nonumber\\
& \quad\times p(\tau(l);a,b)\text{d}\tau(l).\label{n1b}
\end{align}

\section{Iterative marginal maximum likelihood estimator}
The derivation procedure of the IMMLE is presented in this section. To begin with, let $\Lambda$ denote the marginal Log-Likelihood (LL) function, which is obtained from Eq.~(\ref{n1b}), as:
{\small \begin{equation}
\Lambda=\ln p\left(\boldsymbol{z}; \boldsymbol{\xi} \right)
=-LMN\ln\pi-L\ln|\boldsymbol{\Sigma}|+\sum_{l=0}^{L-1}\ln g_{MN}\left(\vecnm{\boldsymbol{\rho}(l)}^2,a,b\right),
\end{equation}}%
in which
{\small \begin{align}
g_{MN}\left(\vecnm{\boldsymbol{\rho}(l)}^2,a,b\right)
=\int_{0}^{+\infty}\frac{\exp\left(-\frac{\vecnm{\boldsymbol{\rho}(l)}^2}{\tau(l)}
\right)}{\tau^{MN}(l)}
p(\tau(l);a,b)\text{d}\tau(l) \nonumber\\
=\left\{
\begin{aligned}
&\frac{2\vecnm{\boldsymbol{\rho}(l)}^{a-MN}K_{a-MN}\left(2\vecnm{\boldsymbol{\rho}(l)}/
b^\frac{1}{2}\right)}{b^\frac{MN+a}{2}\Gamma(a)},\text{K-distr. clutter}, \\
&\frac{b^a\Gamma(MN+a)}{\Gamma(a)\left(\vecnm{\boldsymbol{\rho}(l)}^2+b\right)^{MN+a}},\quad  \text{t-distributed clutter},
\end{aligned}
\right.
\label{g}
\end{align}}
where $K_n(\cdot)$ is the modified Bessel function of the second kind of order $n$ (cf. \citep{tunaley10} for more details).

To begin with, we look for the estimates of the clutter parameters, i.e., of the speckle covariance matrix $\boldsymbol{\Sigma}$, and the texture parameters $a$ and $b$. Let $\hat{\boldsymbol{\Sigma}}$ denote the estimate of $\boldsymbol{\Sigma}$ when all the other unknown parameters are fixed, which can be obtained by solving the equation $\partial\Lambda/\partial\boldsymbol{\Sigma}=0$, as \citep{gini1}:
{\small \begin{equation}\label{Sig_es}
\hat{\boldsymbol{\Sigma}}=\frac{1}{L}\sum_{l=0}^{L-1}h_{MN}\left(\vecnm{\boldsymbol{\rho}(l)}^2,a,b\right)
\cdot\left(\boldsymbol{z}(l)-\boldsymbol{A}\left(\boldsymbol{\theta}\right)\boldsymbol{v}(l)\right)
\left(\boldsymbol{z}(l)-\boldsymbol{A}\left(\boldsymbol{\theta}\right)\boldsymbol{v}(l)\right)^H,
\end{equation}}%
in which
{\small \begin{align}
&h_{MN}\left(\vecnm{\boldsymbol{\rho}(l)}^2,a,b\right)
=-\frac{\frac{\partial g_{MN}\left(\vecnm{\boldsymbol{\rho}(l)}^2,a,b\right)}{\partial \vecnm{\boldsymbol{\rho}(l)}^2}}
{g_{MN}\left(\vecnm{\boldsymbol{\rho}(l)}^2,a,b\right)} \nonumber \\
&=\left\{
\begin{aligned}
&\frac{K_{a-MN-1}\left(2\vecnm{\boldsymbol{\rho}(l)}/
b^\frac{1}{2}\right)}{b^\frac{1}{2}\vecnm{\boldsymbol{\rho}(l)}
K_{a-MN}\left(2\vecnm{\boldsymbol{\rho}(l)}/
b^\frac{1}{2}\right)},\text{K-distributed clutter}, \\
&\frac{MN+a}{\vecnm{\boldsymbol{\rho}(l)}^2+b},\quad \text{t-distributed clutter}.
\end{aligned}
\right.
\label{h}
\end{align}}%
Note that $\hat{\boldsymbol{\Sigma}}$ in Eq.~(\ref{Sig_es}) has an iterative nature, as can be seen from the expression of $\boldsymbol{\rho}(l)$ in Eq.~(\ref{rho}).

We further need to normalize $\hat{\boldsymbol{\Sigma}}$ to fulfill the assumption that $\text{tr}\{\boldsymbol{\Sigma}\}=MN$. Let $\hat{\boldsymbol{\Sigma}}_\text{n}$ denote the normalized estimate $\hat{\boldsymbol{\Sigma}}$, which is:
\begin{equation}\label{Sig_es_n}
\hat{\boldsymbol{\Sigma}}_\text{n}=MN
\frac{\hat{\boldsymbol{\Sigma}}}{\text{tr}\left\{\hat{\boldsymbol{\Sigma}}\right\}}.
\end{equation}

Similarly, the estimates of $a$ and $b$ when other unknown parameters are fixed, denoted by $\hat{a}$ and $\hat{b}$, can be found by equating $\partial\Lambda/\partial a$ and $\partial\Lambda/\partial b$ to zero, respectively, i.e., by solving numerically: 
\begin{equation}
\label{a_es}
\frac{\partial\Lambda}{\partial a}=\sum_{l=0}^{L-1}\frac{j_{MN}\left(\vecnm{\boldsymbol{\rho}(l)}^2,a,b\right)}
{g_{MN}\left(\vecnm{\boldsymbol{\rho}(l)}^2,a,b\right)}=0
\end{equation}
and
\begin{equation}
\label{b_es}
\frac{\partial\Lambda}{\partial b}=\sum_{l=0}^{L-1}\frac{k_{MN}\left(\vecnm{\boldsymbol{\rho}(l)}^2,a,b\right)}
{g_{MN}\left(\vecnm{\boldsymbol{\rho}(l)}^2,a,b\right)}=0,
\end{equation}
w.r.t. $a$ and $b$, respectively, in which
{\small \begin{align}
&j_{MN}\left(\vecnm{\boldsymbol{\rho}(l)}^2,a,b\right)=\frac{\partial g_{MN}\left(\vecnm{\boldsymbol{\rho}(l)}^2,a,b\right)}{\partial a} \nonumber \\
&= \left\{
\begin{aligned}
&-\frac{1}{b^a\Gamma(a)}\int_0^{+\infty}\exp\left(-\frac{\vecnm{\boldsymbol{\rho}(l)}^2}{\tau(l)}
-\frac{\tau(l)}{b}\right)\tau(l)^{-MN+a-1}\\
&\cdot\left(\ln\left(\frac{b}{\tau(l)}\right)+\Psi(a)\right)\text{d}\tau(l),
\quad \text{K-distributed clutter},\\
&-\frac{b^a\Gamma(MN+a)\left(\ln\left(\frac{\vecnm{\boldsymbol{\rho}(l)}^2}{b}+1\right)-\Psi(MN+a)+\Psi(a)\right)}
{\Gamma(MN)\left(\vecnm{\boldsymbol{\rho}(l)}^2+b\right)^{MN+a}},\\
&\text{t-distributed clutter},
\end{aligned}
\right.
\end{align}}
where $\Psi(\cdot)$ denotes the digamma function, and
{\small \begin{align}
&k_{MN}\left(\vecnm{\boldsymbol{\rho}(l)}^2,a,b\right)=\frac{\partial g_{MN}\left(\vecnm{\boldsymbol{\rho}(l)}^2,a,b\right)}{\partial b} \nonumber \\
&=\left\{
\begin{aligned}
&\frac{1}{b^{a+2}\Gamma(a)}\int_0^{+\infty}\exp\left(-\frac{\vecnm{\boldsymbol{\rho}(l)}^2}{\tau(l)}
-\frac{\tau(l)}{b}\right)\\
&\cdot\tau(l)^{-MN+a-1}\cdot\left(\tau(l)-ab\right)\text{d}\tau(l), \text{K-distributed clutter},\\
&-\frac{ab^{a-1}\Gamma(MN+a)\left(-a\vecnm{\boldsymbol{\rho}(l)}^2+MNb\right)}
{\Gamma(a+1)\left(\vecnm{\boldsymbol{\rho}(l)}^2+b\right)^{MN+a+1}},\text{t-distr. clutter}.
\end{aligned}
\right.
\end{align}}%

Next, we consider the estimate $\hat{\boldsymbol{v}}(l)$, by solving $\partial\Lambda/\partial \boldsymbol{v}(l)=0$, which reads
\begin{equation}\label{v_es}
\hat{\boldsymbol{v}}(l)=\left(\tilde{\boldsymbol{A}}^H\left(\boldsymbol{\theta}\right)
\tilde{\boldsymbol{A}}\left(\boldsymbol{\theta}\right)\right)^{-1}
\tilde{\boldsymbol{A}}^H\left(\boldsymbol{\theta}\right)
\tilde{\boldsymbol{z}}(l),
\end{equation}in which
$\tilde{\boldsymbol{A}}\left(\boldsymbol{\theta}\right)=\boldsymbol{\Sigma}^{-\frac{1}{2}}\boldsymbol{A}\left(\boldsymbol{\theta}\right),
$ and
$ \tilde{\boldsymbol{z}}(l)=\boldsymbol{\Sigma}^{-1/2}\boldsymbol{z}(l),$ representing the steering matrix and the observation at pulse $l$, both pre-whitened by the speckle covariance matrix $\boldsymbol{\Sigma}$, respectively.

As the expressions in Eqs.~(\ref{Sig_es}), (\ref{a_es}), (\ref{b_es}) and (\ref{v_es}) suggest, the estimation of each of the parameters $a$, $b$, $\boldsymbol{\Sigma}$ and $\boldsymbol{v}(l)$ requires the knowledge of all the others of them, and furthermore the knowledge of the parameter vector $\boldsymbol{\theta}$. This mutual dependence between the unknown parameters makes it impossible to concentrate the LL function \emph{analytically}, i.e., to obtain a closed-form expression for the LL function concentrated w.r.t. each of the aforementioned parameters that is independent of the other ones. Instead, we resort to the so-called \emph{stepwise numerical concentration} approach.

This approach consists in concentrating the LL function iteratively, by assuming that certain parameters are known from the previous iteration. For the task under consideration, we assume, at each iteration, that $\hat{\boldsymbol{\Sigma}}$, $\hat{a}$ and $\hat{b}$ are known and use them to compute $\hat{\boldsymbol{v}}(l)$, which is then used in turn to update the values of $\hat{\boldsymbol{\Sigma}}$ and $\hat{a}$ and $\hat{b}$ to be used in the next iteration. This sequential updating procedure is repeated until convergence or a maximum iteration number is reached.

Next, we turn to the estimation of $\boldsymbol{\theta}$. The approach explained above allows us to drop all the constant terms in the LL function (including those terms that contain only $\boldsymbol{\Sigma}$, $a$ and $b$ as unknown parameters, as these are assumed to be known at each iteration). Furthermore, by inserting the expression of $\hat{\boldsymbol{v}}(l)$ in Eq.~(\ref{v_es}) into what remains in the LL function, we obtain the estimate of $\boldsymbol{\theta}$, denoted by $\hat{\boldsymbol{\theta}}$, as:
\begin{align}
\hat{\boldsymbol{\theta}}=
\left\{
\begin{aligned}
&\arg\min_{\boldsymbol{\theta}}\Bigg\{\sum_{l=0}^{L-1}\Big((MN-a)\ln\left(\vecnm{
\boldsymbol{P}_{\tilde{\boldsymbol{A}}(\boldsymbol{\theta})}^{\bot}\tilde{\boldsymbol{z}}(l)}\right)\\
& -\ln K_{a-MN}\left(\left.2\vecnm{\boldsymbol{P}_{\tilde{\boldsymbol{A}}(\boldsymbol{\theta})}^{\bot}
\tilde{\boldsymbol{z}}(l)}\middle/
b^\frac{1}{2}\right.\right)\Big)\Bigg\}, \text{K-distr. clutter},\\
&\arg\min_{\boldsymbol{\theta}}\left\{\sum_{l=0}^{L-1}\ln\left(\vecnm{\boldsymbol{P}_{\tilde{\boldsymbol{A}}
(\boldsymbol{\theta})}^{\bot}\tilde{\boldsymbol{z}}(l)}^2+b\right)
\right\}, \text{t-distributed clutter}.
\end{aligned}
\right.
\label{theta_es}
\end{align}in which
$\boldsymbol{P}_{\tilde{\boldsymbol{A}}(\boldsymbol{\theta})}^{\bot}=\boldsymbol{I}_{MN}
-\tilde{\boldsymbol{A}}(\boldsymbol{\theta})\left(\tilde{\boldsymbol{A}}^H(\boldsymbol{\theta})
\tilde{\boldsymbol{A}}(\boldsymbol{\theta})\right)^{-1}\tilde{\boldsymbol{A}}^H(\boldsymbol{\theta})
$ is the orthogonal projection matrix onto the null space of $\tilde{\boldsymbol{A}}(\boldsymbol{\theta})$.
Finally, the whole procedure of the IMMLE is summarized in Table.~1.

\begin{table*}[h!]
 \begin{center}
\begin{tabular}{ll}
\hline \\[-8pt]
\ &  The IMMLE procedures
 \\[2pt]
\hline \\[-8pt]
\ \textbf{Initialization} & $i=0$, set $\hat{a}^{(0)}$, $\hat{b}^{(0)}$ to be two arbitrary positive numbers and $\hat{\boldsymbol{\Sigma}}^{(0)}_\text{n}=\boldsymbol{I}_{MN}$ \\[2pt]
\ \textbf{Step 1} & Iteration $i$, calculate $\hat{\boldsymbol{\theta}}^{(i)}$ from Eq.~(\ref{theta_es}) using $\hat{a}^{(i)}$, $\hat{b}^{(i)}$ and $\hat{\boldsymbol{\Sigma}}^{(i)}_\text{n}$ \\[2pt]
\ & Calculate $\hat{\boldsymbol{v}}^{(i)}(l)$ from Eq.~(\ref{v_es}) using $\hat{\boldsymbol{\theta}}^{(i)}$, $\hat{a}^{(i)}$, $\hat{b}^{(i)}$ and $\hat{\boldsymbol{\Sigma}}^{(i)}_\text{n}$ \\[2pt]
\ \textbf{Step 2} &  Update $\hat{a}^{(i+1)}$ from Eq.~(\ref{a_es}) using  $\hat{\boldsymbol{\theta}}^{(i)}$, $\hat{\boldsymbol{v}}^{(i)}(l)$, $\hat{\boldsymbol{\Sigma}}^{(i)}_\text{n}$, and $\hat{b}^{(i)}$ \\[2pt]
\ &  Update $\hat{b}^{(i+1)}$ from Eq.~(\ref{b_es}) using  $\hat{\boldsymbol{\theta}}^{(i)}$, $\hat{\boldsymbol{v}}^{(i)}(l)$, $\hat{\boldsymbol{\Sigma}}^{(i)}_\text{n}$, and $\hat{a}^{(i+1)}$ \\[2pt]
\ &  Update $\hat{\boldsymbol{\Sigma}}^{(i+1)}_\text{n}$ from Eqs.~(\ref{Sig_es}) and (\ref{Sig_es_n}) using
     $\hat{\boldsymbol{\theta}}^{(i)}$, $\hat{\boldsymbol{v}}^{(i)}(l)$, $\hat{a}^{(i+1)}$ and $\hat{b}^{(i+1)}$\\[2pt]
\ &  Set $i\leftarrow i+1$ \\[2pt]
\ \textbf{Step 3} &  Repeat Step 1 and Step 2 until convergence  \\[2pt]
\hline
\end{tabular}
\caption{Summarization of the proposed algorithm}
 \end{center}
\end{table*}

\noindent\textbf{Remark 1}: Let us recall the expression of $\hat{\boldsymbol{\theta}}$ for the Conventional ML Estimator (CvMLE), which treats the clutter as uniform white Gaussian distributed, denoted by CvMLE-U,
\begin{equation}\label{conv_ml}
\hat{\boldsymbol{\theta}}_{\rm CvMLE-U}= \arg\min_{\boldsymbol{\theta}}\sum_{l=0}^{L-1}\vecnm{\boldsymbol{P}_{{\boldsymbol{A}}
(\boldsymbol{\theta})}^{\bot}{\boldsymbol{z}}(l)}^2,
\end{equation}
as well as for both of the ICdMLE and IJMLE that we proposed in \citep{xzh4,xzh5}, which, adapted to the model in question, has the following expression,
{\small \begin{equation}\label{ICdMLE_ml}
\hat{\boldsymbol{\theta}}_{\rm ICdMLE/IJMLE}=\arg \min_{\boldsymbol{\theta}}\sum_{l=0}^{L-1}\dfrac{1}{\hat{\tau}_{\rm ICdMLE/IJMLE}(l)}
\vecnm{\boldsymbol{P}_{\tilde{\boldsymbol{A}}(\boldsymbol{\theta})}^{\bot}\tilde{\boldsymbol{z}}(l)}^2.
\end{equation}}%
Expression of $\hat{\boldsymbol{\theta}}_{\rm CvMLE-U}$ shows that the CvMLE-U considers simply the \emph{sum} of $\vecnm{\boldsymbol{P}_{{\boldsymbol{A}}
(\boldsymbol{\theta})}^{\bot}{\boldsymbol{z}}(l)}^2$ (the square of the norm of the projection of the observation at pulse $l$ onto the null space of the steering matrix), while the ICdMLE and IJMLE, as the expression of $\hat{\boldsymbol{\theta}}_{\rm ICdMLE/IJMLE}$ shows, consider the \emph{modified sum} of these terms (pre-whitened by the speckle covariance matrix, and weighted by the inverse of the texture realization at each pulse). It is precisely because of this modification that the ICdMLE and IJMLE gain their advantages in performance over the CvMLE-U. An iterative version of the CvMLE-U with no assumption on the covariance matrix, denoted by ICvMLE, can be easily derived, for which, the aforementioned conclusions remain valid for the ICvMLE. On the other hand, we can see from Eq.~(\ref{theta_es}) that the proposed IMMLE considers, instead of direct or modified sum of the projections, the sum of their \emph{logarithms} (modified by some algebraic operations), which is equivalent to the \emph{product} of them. Since a sum is small only if all its terms are small, while a product can be small even if only very few of its terms are small enough, we can conclude that underlying this contrast between summation and multiplication is a difference \textit{in essentia}, that the CvMLE-U, ICvMLE, ICdMLE and IJMLE treat all the pulses ``equally'', whereas the IMMLE focuses only on the ``best'' pulses. Due to space limitation, refer to Table. 2 for a concise comparison between the IMMLE, IJMLE and ICdMLE.

\begin{table}[h!]
 \begin{center}
{\small
\begin{tabular}{|m{3cm}|m{1.6cm}|m{1.3cm}|m{1cm}|}
\hline
\ & ICdMLE & IJMLE & IMMLE\\
\hline
Likelihood & Conditional & Joint & Marginal\\
\hline
Texture modeling & Deterministic & \multicolumn{2}{c|}{Stochastic}\\
\hline
Considers $\boldsymbol{\tau}$ & \multicolumn{2}{c|}{Yes} & No\\
\hline
Considers $a$ and $b$ & No & \multicolumn{2}{c|}{Yes}\\
\hline
Numerical solution of equations & {No} & \multicolumn{2}{c|}{{Yes}}\\
\hline
Numerical integration & \multicolumn{2}{c|}{{No}} & {Yes}\\
\hline
Computational complexity & {Lowest} & {Higher than ICdMLE} & {Highest}\\
\hline
Iteration(s) required & \multicolumn{2}{c|}{{Two}} & {One}\\
\hline
Requires texture distribution & {No} & \multicolumn{2}{c|}{{Yes}}\\
\hline
Can be used for texture parameters estimation & {No} & \multicolumn{2}{c|}{{Yes}} \\
\hline
\end{tabular}}%
\caption{Comparison between ICdMLE, IJMLE and IMMLE }
 \end{center}
\end{table}

\noindent\textbf{Remark 2}: As is clear from the procedure above, our algorithm does not entail the estimation of the RCS coefficients $\alpha_k$, and the normalized Doppler frequencies $f_k$, of the targets, but rather only involves estimating the vectors $\boldsymbol{v}(l)$, which are functions of them. Indeed, in applications where the estimation of those parameters are of interest, one can naturally find the ML or LS estimates of them by respectively equating an adequate cost function to zero, and then complement our algorithm accordingly. This, however, deviates from our topic, i.e., the DOD/DOA estimation, and due to space limitation, it is not to be discussed in this paper.

\noindent\textbf{Remark 3}: The convergence of the LL function is guaranteed by the fact that the value of the objective function to calculate $\hat{\boldsymbol{\theta}}$, $\hat{\boldsymbol{\Sigma}}$, $\hat{a}$ and $\hat{b}$ at each step can either improve or maintain but cannot worsen. As the simulations will show, the convergence of the estimates of the unknown parameters in $\boldsymbol{\theta}$ can be obtained by few iterations (one to two).

\noindent\textbf{Remark 4}: IMMLE has a computational complexity sightly higher than ICvMLE, ML-GM, IJMLE and CdMLE. Indeed, all of them possess a highly non-convex minimization step over a $2K$-dimensional parameter space, which is the most time-consuming stage compared to the updating steps of the speckle covariance matrix, $\boldsymbol{\Sigma}$ and the vector $\boldsymbol{v}$ (both of them mainly based on analytical expressions) and the potential numerical solving. Generally, a MUSIC-based algorithm has a lower complexity than the ML-based one, except for $\ell_p$-MUSIC algorithm where the signal/noise subspaces construction is time-consuming due to the $\ell_p$ norm minimization.

\section{Cram\'{e}r-Rao bound expression}
The CRB w.r.t. target direction parameters in a MIMO radar context in the presence of SIRP clutter has been derived in our previous works \citep{xzh3}, where we used an element-wise approach to calculate the Fisher information matrix (FIM). For the model considered in this paper, where the size of the unknown signal parameter vector (hence the dimension of the resulting FIM) is much larger, a block-wise expression for the CRB w.r.t. the signal DODs and DOAs (denoted by $\text{CRB}\left(\boldsymbol{\theta}\right)$) is required, the result of which is presented below. 
 The $2K\times 2K$ CRB matrix w.r.t. $\boldsymbol{\theta}$ in the presence of SIRP clutter is given by:
\begin{align}
\text{CRB}\left(\boldsymbol{\theta}\right)&=\left(\frac{2\kappa}{MN}\Re\left\{\sum_{l=0}^{L-1}\boldsymbol{H}^H(l)\tilde{\boldsymbol{D}}^H
\boldsymbol{P}_{\tilde{\boldsymbol{A}}(\boldsymbol{\theta})}^{\bot}\tilde{\boldsymbol{D}}\boldsymbol{H}(l)\right\}\right)^{-1} \nonumber \\
&=\frac{MN}{2\kappa L}\left(\Re\left\{\left(\tilde{\boldsymbol{D}}^H\boldsymbol{P}_{\tilde{\boldsymbol{A}}(\boldsymbol{\theta})}^{\bot}
\tilde{\boldsymbol{D}}\right)\odot\hat{\boldsymbol{P}}^T\right\}\right)^{-1},
\label{crb}
\end{align}in which
\begin{equation}\label{kappa}
\kappa=\left\{
\begin{aligned}
&\frac{\int_0^{+\infty}x^{MN+a-1}\frac{K_{a-MN-1}^2(x)}
{K_{a-MN}(x)}\text{d}x}{2^{MN+a-2}b\Gamma(MN)\Gamma(a)}, \quad \text{K-distributed clutter}, \\
&\frac{MNa(a+MN)}{b(a+MN+1)},\quad \text{t-distributed clutter},
\end{aligned}
\right.
\end{equation}where $K_n(x)$ is the modified Bessel functions of the second kind of order $n$, $\boldsymbol{H}(l)=\boldsymbol{I}_2\otimes\text{diag}\left\{\left[\boldsymbol{v}(l)\right]_1,\dots,\left[\boldsymbol{v}(l)\right]_K\right\}$, $\boldsymbol{J}_2$ is the all-ones matrix of size $2$, 
$\hat{\boldsymbol{P}}=\frac{1}{L}\boldsymbol{J}_2\otimes\sum_{l=0}^{L-1}\boldsymbol{v}(l)\boldsymbol{v}^H(l),$ and
{\small \begin{align*}
&\hspace*{2.5cm}\tilde{\boldsymbol{D}}=\boldsymbol{\Sigma}^{-\frac{1}{2}}\left[\boldsymbol{D}^{(\mathcal T)},\boldsymbol{D}^{(\mathcal R)}\right]\\
&\text{where}\,\left\lbrace\begin{array}{l}
\boldsymbol{D}^{(\mathcal T)}=\left[\left.\frac{\partial\boldsymbol{a}\left(\theta^{(\mathcal T)},\theta^{(\mathcal R)}\right)}
{\partial\theta^{(\mathcal T)}}\right|_{\theta^{(\mathcal T)}=\theta^{(\mathcal T)}_1,\theta^{(\mathcal R)}=\theta^{(\mathcal R)}_1},
\dots,\right.\\
\hspace*{2cm}\left.\left.\frac{\partial\boldsymbol{a}\left(\theta^{(\mathcal T)},\theta^{(\mathcal R)}\right)}
{\partial\theta^{(\mathcal T)}}\right|_{\theta^{(\mathcal T)}=\theta^{(\mathcal T)}_K,\theta^{(\mathcal R)}=\theta^{(\mathcal R)}_K}\right] \\
\\
\boldsymbol{D}^{(\mathcal R)}=\left[
\left.\frac{\partial\boldsymbol{a}\left(\theta^{(\mathcal T)},\theta^{(\mathcal R)}\right)}
{\partial\theta^{(\mathcal R)}}\right|_{\theta^{(\mathcal T)}=\theta^{(\mathcal T)}_1,\theta^{(\mathcal R)}=\theta^{(\mathcal R)}_1},
\dots,\right.\\
\hspace*{2cm}\left.\left.\frac{\partial\boldsymbol{a}\left(\theta^{(\mathcal T)},\theta^{(\mathcal R)}\right)}
{\partial\theta^{(\mathcal R)}}\right|_{\theta^{(\mathcal T)}=\theta^{(\mathcal T)}_K,\theta^{(\mathcal R)}=\theta^{(\mathcal R)}_K}\right]
\end{array}\right.
\end{align*}}

\section{Numerical simulations}
For simulations, we consider a MIMO radar comprising $M=3$ sensors at the transmitter and $N=4$ at the receiver, both with half-wave length inter-element spacing. The DOD and DOA of the first source are respectively $18^\circ$ and $20^\circ$, and of the second source are $45^\circ$ and $40^\circ$. The coefficients $\alpha_1$ and $\alpha_2$ are chosen to be $2+3j$ and $1-0.5j$, and the normalized Doppler frequencies $f_1$ and $f_2$ are $0.3$ and $0.8$. There are $L=15$ pulses per CPI, and each pulse contains $T=5$ snapshots. For K-distributed clutter, we choose $a=2$ and $b=10$; and for t-distributed clutter, $a=1.1$ and $b=2$. The entries of the speckle covariance matrix $\boldsymbol{\Sigma}$ are generated by $[\boldsymbol{\Sigma}]_{m,n}=\sigma^2 0.9^{|m-n|}e^{j\frac{\pi}{2}(m-n)}, \ m,n=1,\dots,MN$, in which $\sigma^2$ is a factor to adjust speckle power. Each point of the MSE in the figures is generated by averaging the results of $500$ Monte-Carlo trials. The signal-to-clutter ratio (SCR) \citep{Akcakaya1} is defined by
$
\text{SCR}=\frac{1}{L}\frac{\sum_{l=0}^{L-1}\left(\boldsymbol{A}\left(\boldsymbol{\theta}\right)\boldsymbol{v}(l)\right)^H
\left(\boldsymbol{A}\left(\boldsymbol{\theta}\right)\boldsymbol{v}(l)\right)}
{\text{E}\{\tau(l)\}\text{tr}\left\{\boldsymbol{\Sigma}\right\}},
$
in which $\text{E}\{\tau(l)\}$ is equal to $ab$ for a K-distributed clutter and $b/(a-1)$ for a t-distributed clutter (for $a>1$).

\begin{figure*}[h!]
	\centering
%
\definecolor{mycolor1}{rgb}{0.00000,0.44700,0.74100}%
\definecolor{mycolor2}{rgb}{0.85000,0.32500,0.09800}%
\definecolor{mycolor3}{rgb}{0.92900,0.69400,0.12500}%
\definecolor{mycolor4}{rgb}{0.49400,0.18400,0.55600}%
\definecolor{mycolor5}{rgb}{0.46600,0.67400,0.18800}%
\definecolor{mycolor6}{rgb}{0.30100,0.74500,0.93300}%
\definecolor{mycolor7}{rgb}{0.63500,0.07800,0.18400}%
\begin{tikzpicture}

\begin{axis}[%
width=0.65\linewidth,
height=0.45\linewidth,
at={(0,0)},
scale only axis,
xmin=-5,
xmax=30,
xlabel style={font=\color{white!15!black}},
xlabel={SCR (dB)},
ymin=-25,
ymax=37,
ylabel style={font=\color{white!15!black}},
ylabel={MSE (dB)},
axis background/.style={fill=white},
xmajorgrids,
ymajorgrids,
legend columns=2,
legend style={at={(0.01,0.01)}, anchor=south west, legend cell align=left, align=left, font=\footnotesize, draw=white!15!black}
]
\addplot [color=black,line width=1.0pt]
  table[row sep=crcr]{%
-5	17.2508900329105\\
0	12.2508900329105\\
5	7.2508900329105\\
10	2.2508900329105\\
15	-2.7491099670895\\
20	-7.7491099670895\\
25	-12.7491099670895\\
30	-17.7491099670895\\
};
\addlegendentry{CRB}

\addplot [color=mycolor1, mark=+, mark size=3pt, mark options={solid, mycolor1}]
  table[row sep=crcr]{%
-5	32.3598308555755\\
0	30.7718812229055\\
5	30.074207663765\\
10	23.0847414220099\\
15	4.62201057766454\\
20	-1.06808484113982\\
25	-6.12841803008099\\
30	-11.2958940468951\\
};
\addlegendentry{MUSIC-SCM}

\addplot [color=mycolor2, mark=o, mark options={solid, mycolor2}]
  table[row sep=crcr]{%
-5	31.913195748139\\
0	30.7067320710952\\
5	28.8025588029622\\
10	16.1937512095654\\
15	2.80276644685539\\
20	-2.26434940545095\\
25	-7.20757221542332\\
30	-12.1838192830026\\
};
\addlegendentry{MUSIC-Tyler}

\addplot [color=mycolor3, mark=asterisk, mark size=3pt, mark options={solid, mycolor3}]
  table[row sep=crcr]{%
-5	32.0940440131789\\
0	30.6504285735365\\
5	29.4276124948098\\
10	18.7768983452929\\
15	2.91890552426\\
20	-2.33223589278398\\
25	-7.33396431339987\\
30	-12.4689918735972\\
};
\addlegendentry{RG-MUSIC}

\addplot [color=mycolor4, mark=square, mark options={solid, mycolor4}]
  table[row sep=crcr]{%
-5	32.2133941865729\\
0	30.9522970630482\\
5	29.0840295232914\\
10	19.9920894641884\\
15	6.13188476300863\\
20	0.692337467445948\\
25	-4.19814187418907\\
30	-8.77619469342471\\
};
\addlegendentry{MKG algorithm}

\addplot [color=mycolor5, mark=x, mark size=3pt, mark options={solid, mycolor5}]
  table[row sep=crcr]{%
-5	31.7545533217921\\
0	30.7889009917191\\
5	30.0189019127382\\
10	22.6398024326773\\
15	3.78381917848632\\
20	-1.16660310223472\\
25	-6.16140533472177\\
30	-11.3480667130031\\
};
\addlegendentry{ML-GM}


\addplot [color=mycolor7, mark=triangle, mark size=3pt, mark options={solid, rotate=180, mycolor7}]
  table[row sep=crcr]{%
-5	33.3418171298028\\
0	32.4093009738474\\
5	24.6390718786166\\
10	10.1500928391609\\
15	3.9907557805058\\
20	-1.18890330191206\\
25	-6.15645027935796\\
30	-11.3110902631891\\
};
\addlegendentry{ICvMLE 1 iteration}

\addplot [color=mycolor1, dashdotted, mark=triangle, mark size=3pt, mark options={solid, rotate=180, mycolor1}]
  table[row sep=crcr]{%
-5	33.1533574292897\\
0	32.2511500871099\\
5	24.2302325635749\\
10	10.1500992451213\\
15	3.99076112368137\\
20	-1.18893661577739\\
25	-6.1564592740419\\
30	-11.3111002420417\\
};
\addlegendentry{ICvMLE 2 iterations}

\addplot [color=mycolor2, mark=star, mark size=3pt, mark options={solid, mycolor2}]
  table[row sep=crcr]{%
-5	33.0439900144275\\
0	30.3803874757076\\
5	16.0036766359803\\
10	7.50732415372852\\
15	1.43848478396779\\
20	-3.56494081921487\\
25	-8.48235745867095\\
30	-13.8357699932039\\
};
\addlegendentry{ICdMLE 1 iteration}

\addplot [color=mycolor3, dashdotted, mark=star, mark size=3pt, mark options={solid, mycolor3}]
  table[row sep=crcr]{%
-5	32.8438092760062\\
0	29.3861326041991\\
5	14.1722116329548\\
10	6.23577049921356\\
15	0.413246663425276\\
20	-4.43955299174062\\
25	-9.32371842390668\\
30	-14.7657747849965\\
};
\addlegendentry{ICdMLE 2 iterations}

\addplot [color=mycolor4, mark=triangle, mark size=3pt, mark options={solid, mycolor4}]
  table[row sep=crcr]{%
-5	32.9812034725008\\
0	30.246187411067\\
5	15.720041524083\\
10	7.50741705015466\\
15	1.43856832235116\\
20	-3.56473100727341\\
25	-8.48278086119254\\
30	-13.8354803611086\\
};
\addlegendentry{IJMLE 1 iteration}

\addplot [color=mycolor5, dashdotted, mark=triangle, mark size=3pt, mark options={solid, mycolor5}]
  table[row sep=crcr]{%
-5	32.6757999048057\\
0	29.3705530482794\\
5	15.202655983496\\
10	6.23920900165684\\
15	0.436968464200302\\
20	-4.42019942216168\\
25	-9.31685574546806\\
30	-14.76010345287\\
};
\addlegendentry{IJMLE 2 iterations}

\addplot [color=mycolor6, mark=diamond, mark size=3pt, mark options={solid, mycolor6}]
  table[row sep=crcr]{%
-5	32.5442279130978\\
0	26.64159011689\\
5	11.3292022662137\\
10	5.59319903131408\\
15	-0.240692680919402\\
20	-4.59480986110889\\
25	-9.9553174493905\\
30	-14.7856371642414\\
};
\addlegendentry{IMMLE 1 iteration}

\addplot [color=mycolor7, dashdotted, mark=diamond, mark size=3pt, mark options={solid, mycolor7}]
  table[row sep=crcr]{%
-5	32.4844454637996\\
0	26.1256393765894\\
5	10.7891678439724\\
10	5.23434250986144\\
15	-0.495188751266113\\
20	-4.91704047569308\\
25	-10.0544408328028\\
30	-15.1503680441016\\
};
\addlegendentry{IMMLE 2 iterations}

\addplot [color=mycolor1, mark=triangle, mark size=3pt, mark options={solid, rotate=270, mycolor1}]
  table[row sep=crcr]{%
-5	32.2689550863927\\
0	30.722775297772\\
5	29.9142484346259\\
10	20.9210046765351\\
15	3.62958601235799\\
20	-1.72613200277334\\
25	-6.84554652930174\\
30	-12.1179696670678\\
};
\addlegendentry{$\ell_\text{p}\text{-MUSIC, p = 1.20}$}

\addplot [color=mycolor2, mark=triangle, mark size=3pt, mark options={solid, rotate = 90, mycolor2}]
  table[row sep=crcr]{%
-5	32.4473714455163\\
0	30.7888973307046\\
5	30.084574060214\\
10	23.0183631601361\\
15	4.59549663239739\\
20	-1.08091143410075\\
25	-6.12483077576059\\
30	-11.2909768686244\\
};
\addlegendentry{ROC-MUSIC}

\end{axis}
\end{tikzpicture}%
      \caption{MSE vs. SCR under K-distributed clutter, L = $15$}
	\label{MSE_vs_SCR_Kdist}
\end{figure*}
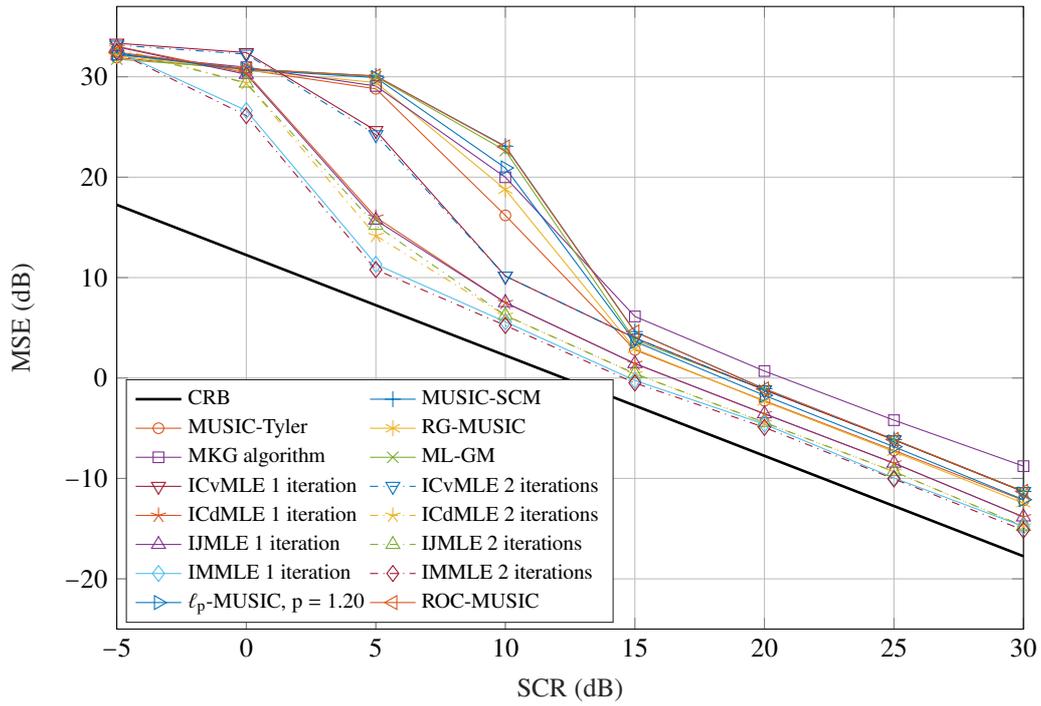

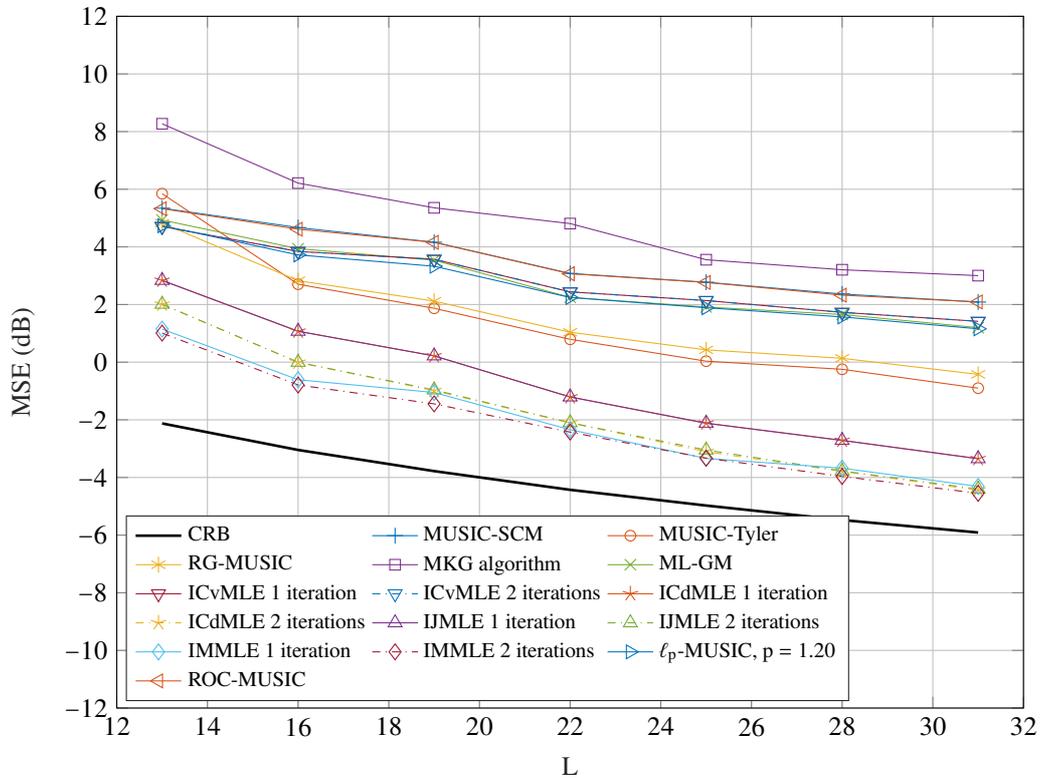
\begin{figure*}[h!]
	\centering
%
\definecolor{mycolor1}{rgb}{0.00000,0.44700,0.74100}%
\definecolor{mycolor2}{rgb}{0.85000,0.32500,0.09800}%
\definecolor{mycolor3}{rgb}{0.92900,0.69400,0.12500}%
\definecolor{mycolor4}{rgb}{0.49400,0.18400,0.55600}%
\definecolor{mycolor5}{rgb}{0.46600,0.67400,0.18800}%
\definecolor{mycolor6}{rgb}{0.30100,0.74500,0.93300}%
\definecolor{mycolor7}{rgb}{0.63500,0.07800,0.18400}%
\begin{tikzpicture}

\begin{axis}[%
width=0.65\linewidth,
height=0.5\linewidth,
at={(0,0)},
scale only axis,
xmin=12,
xmax=32,
xlabel style={font=\color{white!15!black}},
xlabel={L},
ymin=-12,
ymax=12,
ylabel style={font=\color{white!15!black}},
ylabel={MSE (dB)},
axis background/.style={fill=white},
xmajorgrids,
ymajorgrids,
legend columns=3,
legend style={at={(0.01,0.01)}, anchor=south west, legend cell align=left, font=\footnotesize, align=left, draw=white!15!black}
]
\addplot [color=black,line width=1.0pt]
  table[row sep=crcr]{%
13	-2.12503665924496\\
16	-3.04573487437271\\
19	-3.77924848910325\\
22	-4.42876185603554\\
25	-4.97424167918363\\
28	-5.47611536123566\\
31	-5.91036023282292\\
};
\addlegendentry{CRB}

\addplot [color=mycolor1, mark=+, mark size=3pt, mark options={solid, mycolor1}]
  table[row sep=crcr]{%
13	5.34200865027917\\
16	4.67029347708101\\
19	4.16180998047081\\
22	3.07816930261003\\
25	2.77403791787385\\
28	2.36545965816967\\
31	2.08677452585518\\
};
\addlegendentry{MUSIC-SCM}

\addplot [color=mycolor2, mark=o, mark options={solid, mycolor2}]
  table[row sep=crcr]{%
13	5.84583573661611\\
16	2.70390742068776\\
19	1.86832630990415\\
22	0.792666040283407\\
25	0.034078918747269\\
28	-0.24720296792661\\
31	-0.902045293613555\\
};
\addlegendentry{MUSIC-Tyler}

\addplot [color=mycolor3, mark=asterisk, mark size=3pt, mark options={solid, mycolor3}]
  table[row sep=crcr]{%
13	4.82108832368481\\
16	2.8268637329063\\
19	2.11805038058914\\
22	1.03883624577525\\
25	0.426947886969083\\
28	0.134347093056117\\
31	-0.423624808100614\\
};
\addlegendentry{RG-MUSIC}

\addplot [color=mycolor4, mark=square, mark options={solid, mycolor4}]
  table[row sep=crcr]{%
13	8.27084331117566\\
16	6.21253675808932\\
19	5.35528227971002\\
22	4.80858345077382\\
25	3.55422731389135\\
28	3.20638887432316\\
31	3.00236161508596\\
};
\addlegendentry{MKG algorithm}

\addplot [color=mycolor5, mark=x, mark size=3pt, mark options={solid, mycolor5}]
  table[row sep=crcr]{%
13	4.9315139580955\\
16	3.93795909419758\\
19	3.53827362012646\\
22	2.24893313075016\\
25	1.91384334001585\\
28	1.6531899391321\\
31	1.19699471947517\\
};
\addlegendentry{ML-GM}


\addplot [color=mycolor7, mark=triangle, mark size=3pt, mark options={solid, rotate=180, mycolor7}]
  table[row sep=crcr]{%
13	4.70690244808677\\
16	3.8419125609967\\
19	3.56994699970741\\
22	2.43862557368277\\
25	2.1385993045444\\
28	1.72848320618457\\
31	1.41604530857497\\
};
\addlegendentry{ICvMLE 1 iteration}

\addplot [color=mycolor1, dashdotted, mark=triangle, mark size=3pt, mark options={solid, rotate=180, mycolor1}]
  table[row sep=crcr]{%
13	4.70690315492662\\
16	3.8419311642665\\
19	3.56991029319804\\
22	2.43862797735964\\
25	2.13860009172997\\
28	1.72848952801925\\
31	1.41603823658067\\
};
\addlegendentry{ICvMLE 2 iterations}

\addplot [color=mycolor2, mark=star, mark size=3pt, mark options={solid, mycolor2}]
  table[row sep=crcr]{%
13	2.8370723556996\\
16	1.07268656456507\\
19	0.217775165685452\\
22	-1.2066879360094\\
25	-2.11727583302932\\
28	-2.71345694696656\\
31	-3.3479177497554\\
};
\addlegendentry{ICdMLE 1 iteration}

\addplot [color=mycolor3, dashdotted, mark=star, mark size=3pt, mark options={solid, mycolor3}]
  table[row sep=crcr]{%
13	2.0099269663963\\
16	-0.00564236840740202\\
19	-0.988170643268263\\
22	-2.10595386297369\\
25	-3.10534859826046\\
28	-3.79025027159806\\
31	-4.43961304922959\\
};
\addlegendentry{ICdMLE 2 iterations}

\addplot [color=mycolor4, mark=triangle, mark size=3pt, mark options={solid, mycolor4}]
  table[row sep=crcr]{%
13	2.83715008270573\\
16	1.07257586015617\\
19	0.218033736280269\\
22	-1.20676772994257\\
25	-2.11693166543054\\
28	-2.71317950175459\\
31	-3.34783731031251\\
};
\addlegendentry{IJMLE 1 iteration}

\addplot [color=mycolor5, dashdotted, mark=triangle, mark size=3pt, mark options={solid, mycolor5}]
  table[row sep=crcr]{%
13	2.01004724474715\\
16	-0.0033725216640779\\
19	-0.960174275217844\\
22	-2.10071142500053\\
25	-3.04645974693308\\
28	-3.77401445308171\\
31	-4.40088551025299\\
};
\addlegendentry{IJMLE 2 iterations}

\addplot [color=mycolor6, mark=diamond, mark size=3pt, mark options={solid, mycolor6}]
  table[row sep=crcr]{%
13	1.14856055269019\\
16	-0.609234009810277\\
19	-1.05172049642272\\
22	-2.34258651095271\\
25	-3.33860403106655\\
28	-3.67805227048944\\
31	-4.31090491115166\\
};
\addlegendentry{IMMLE 1 iteration}

\addplot [color=mycolor7, dashdotted, mark=diamond, mark size=3pt, mark options={solid, mycolor7}]
  table[row sep=crcr]{%
13	1.00976518920751\\
16	-0.79151114146474\\
19	-1.44629852239809\\
22	-2.42628384407796\\
25	-3.32902536781181\\
28	-3.96338954479133\\
31	-4.54873521535741\\
};
\addlegendentry{IMMLE 2 iterations}

\addplot [color=mycolor1, mark=triangle, mark size=3pt, mark options={solid, rotate=270, mycolor1}]
  table[row sep=crcr]{%
13	4.73528199356049\\
16	3.72167200456332\\
19	3.32804206802355\\
22	2.2450887336792\\
25	1.8881946485923\\
28	1.57034578095558\\
31	1.15770123571542\\
};
\addlegendentry{$\ell_\text{p}\text{-MUSIC, p = 1.20}$}

\addplot [color=mycolor2, mark=triangle, mark size=3pt, mark options={solid, rotate = 90, mycolor2}]
  table[row sep=crcr]{%
13	5.32445706428739\\
16	4.61268181404751\\
19	4.15744034328384\\
22	3.06753185089526\\
25	2.76703163566032\\
28	2.32524959053121\\
31	2.08900370051012\\
};
\addlegendentry{ROC-MUSIC}

\end{axis}
\end{tikzpicture}%
      \caption{MSE vs. $L$ under K-distributed clutter, SCR = $15$ dB}
	\label{MSE_vs_L_Kdist}
\end{figure*}

Figs.~\ref{MSE_vs_SCR_Kdist} and \ref{MSE_vs_L_Kdist} investigate the performance of the proposed IMMLE estimator compared to the classical MUSIC method based on the Sample Covariance Matrix (MUSIC-SCM) as well, its robust version based on the well-known Tyler estimate of the covariance matrix \citep{Tyl87} (MUSIC-Tyler). Others robust MUSIC based algorithms are also considered such as the RG-MUSIC \citep{Cou15}, the $\ell_p$-MUSIC \citep{ZSH13} and the ROC-MUSIC \citep{TN96} as well the MKG algorithm proposed in \citep{BAJ16} and the ML-GM \citep{KS00}. Finally, we consider the ICdMLE, the IJMLE and the ICvMLE, as well the derived CRB. 

In Fig.~\ref{MSE_vs_SCR_Kdist} and \ref{MSE_vs_L_Kdist}, the MSEs  are plotted versus SCR with a fixed $L$, respectively versus the pulse number $L$ with a fixed SCR. \newline
It can be noticed that MUSIC-based algorithms, even the robust versions, do not outperform the proposed algorithm due to a
small number of pulses, which is a typical scenario in radar application. The MKG algorithm assumes a mixture of K-distributed and Gaussian noise, both of them, with a covariance matrix equals to the identity, which explains its poor performance. Whereas, the robust ML-GM is based on, empirically defined number of, Gaussian mixture with identity covariance matrix assumptions. Since, it is a ML estimator (i.e., an estimator based on a parametric model), its accuracy deteriorates if we deviate from the assumed model distribution. Concerning the ICdMLE, IJMLE and the ICvMLE, their performances are below the proposed algorithm. Consequently, from Figures \ref{MSE_vs_SCR_Kdist} et \ref{MSE_vs_L_Kdist}, we can assess that the IMMLE outperforms the aforementioned algorithms. The same behavior is noticed under the t-distributed clutter whether it is MSE versus SCR with fixed $L$ or versus $L$ with fixed SCR.
Finally, the reader is referred to Table. 2 for a concise comparison between the IMMLE, IJMLE and ICdMLE.
\newline
\noindent \textbf{Remark 5:} It is worth mentioning that, the proposed IMMLE estimates are approximation of the true ML estimates due to the iterative stepwise procedure, in which we have to solve numerically at each step three equations for the update of the parameters of the texture distribution and the speckle covariance matrix. Thus these latter, are not exact solutions either. Furthermore, it is worth mentioning that a theoretical analysis of the efficiency of the estimator on $\boldsymbol{\theta}$ is beyond the scope of this paper. Nevertheless, from our extensive simulations, we believe that our proposed algorithm would not be  statistically efficient (i.e., its MSE does not attain the CRB).
\section{Conclusion}
This paper is dedicated to the design of the exact ML DOD and DOA estimation for MIMO radar in the presence of SIRP clutter. Specifically, our proposed iterative estimator is based on the marginal likelihood for which its related cost function is solved using stepwise numerical concentration approach. Finally, interconnections with the existing based likelihood methods, namely, the conventional, the conditional and the joint likelihood based estimators are investigated theoretically and numerically. 

\bibliographystyle{elsarticle-num}
\bibliography{zhxtc}

\end{document}